\documentclass[iop]{emulateapj}

\usepackage{amsmath, amssymb}

\usepackage{natbib}

\usepackage{aas_macros}

\usepackage[T1]{fontenc}

\newcommand{\Brms}{{B_{rms}^{}}}

\defcitealias{Fabbian+2010}{I}
\defcitealias{Fabbian+2012}{II}
\defcitealias{Pereira+2013a}{III}

\shorttitle{Continuum CLV and [O\,\textsc{i}] spectral lines in
  solar 3D (M)HD models}

\begin{document}

\title{Continuum intensity and [O\,\textsc{i}] spectral line profiles in solar 3D photospheric models: \\
  the effect of magnetic fields
}

\author{D. Fabbian} \affil{Instituto de Astrof\'{i}sica de Canarias
  (IAC), E-38205 La Laguna, Tenerife, Spain; damian@iac.es}

\and

\author{F. Moreno-Insertis\altaffilmark{1}} \affil{Instituto de
  Astrof\'{i}sica de Canarias (IAC), E-38205 La Laguna, Tenerife,
  Spain; fmi@iac.es}

\altaffiltext{1}{Also affiliated with: Departamento de
  Astrof\'{i}sica, Universidad de La Laguna (ULL), E-38206 La Laguna,
  Tenerife, Spain}

\begin{abstract}

  The importance of magnetic fields in three-dimensional
  magnetoconvection models of the Sun's photosphere is investigated in
  terms of their influence on the continuum intensity 
    at different viewing inclination angles, and on the
  intensity profile of two [O\,\textsc{i}] spectral lines. We use the
  \textsc{RH} numerical radiative transfer code to perform \textit{a
    posteriori} spectral synthesis on the same time-series of
  magnetoconvection models used in our publications on the
  effect of magnetic fields on abundance determination. We
  obtain a good match of the synthetic disc-centre 
  continuum intensity to the absolute continuum values from the FTS
  observational spectrum; the match of the centre-to-limb variation
  (CLV) synthetic data to observations is also good,
  thanks, in part, to the 3D radiation transfer capabilities of the
  \textsc{RH} code. The different levels of magnetic flux in the
  numerical time-series do not modify the quality of the match. 
  Concerning the targetted [O\,\textsc{i}] spectral lines, 
  we find, instead,
  that magnetic fields lead to non-negligible changes
  in the synthetic spectrum, with larger
  average magnetic flux causing the line to become noticeably weaker.
  The photospheric oxygen abundance that one would derive if
    instead using non-magnetic numerical models would thus be lower by
    a few to several centidexes. The inclusion of
  magnetic fields is confirmed to be important for improving
    the current modelling of the Sun, here in particular in terms of
  spectral line formation and of deriving consistent chemical
    abundances. These results may shed further light on the
  still controversial issue regarding the precise value of the solar
  oxygen abundance.

\end{abstract}


\section{Introduction}\label{sec:intro}

The Sun is magnetised over its whole surface, even in the so called
``quiet'' regions (e.~g.,
\citealt{Trujillo-BuenoShchukinaAsensio-Ramos2004, Solanki2009,
  Sanchez-AlmeidaMartinez-Gonzalez2011, Bellot-RubioOrozco-Suarez2012,
  NordlundSteinAsplund2009}), with at least as much magnetic flux present 
in the quiet Sun itself as in active regions (\citealp{Gosic+2014, 
ThorntonParnell2011}). The presence of ubiquitous magnetic fields
has an influence on the average temperature stratification in the
solar photosphere, thus indirectly affecting the many spectral lines
formed there (see \citealt{Fabbian+2010,Fabbian+2012}, called Papers I
and II, respectively, in the following). This, in turn, can have an
impact on the accuracy of the chemical composition derived from
different spectral features, when non-magnetic atmospheric models,
even current generation ones accounting for departures from
homogeneity
\citep[e.g.,][]{SteinNordlund1998,Asplund+2000,Freytag+2002,
  Asplund2005, Scott+2014}, are employed for solar abundance
derivation purposes. 

Three-dimensional (3D) convection simulations have already been
extensively applied to aid in the understanding of different solar
phenomena, including the impact of the convective velocity field on
the formation of spectral lines in terms of their widths, shifts,
shapes and asymmetries \citep[e.~g.,][]{Dravins+1981, Dravins1982,
  NordlundSteinAsplund2009}. Generally, though,
the effects of magnetic fields on the determination of solar chemical
abundances has not been systematically explored.
In Papers \citetalias{Fabbian+2010} and \citetalias{Fabbian+2012}, we
carried out a first quantitative assessment of the impact of magnetic
fields in 3D convection models on the determination of the solar
chemical composition. To that end, we calculated time-series of
magnetoconvection simulations using the Copenhagen \textsc{STAGGER}
radiation magnetohydrodynamics code.  The quality of those simulations
was confirmed in follow-up studies, where the snapshots were tested
for thermodynamical and polarisation properties
\citep{Beck+2013,Beck+2015a}, as well as being employed for
associating G-band bright points with magnetic field strengths of a
few kG \citep{CriscuoliUitenbroek2014a}, and for comparing
numerically-predicted and observationally-derived velocity fields in
solar convection via Fourier power spectra
\citep{Yelles-ChaoucheMoreno-InsertisBonet+2014}. Our 3D
magnetohydrodynamical (MHD) models were also used for the
interpretation of solar irradiance measurements, whereby the
predictions of the simulations may provide a possible explanation for
the counter-phase variation of the spectral irradiance at visible and
IR wavelengths with the solar activity cycle
\citep{CriscuoliUitenbroek2014b}.

The motivation for the present paper is to further test the quality of
our magnetoconvection models and to proceed in the investigation of
the effect of magnetic fields on the average spectra and derived
chemical composition of the solar photosphere. Using the MHD
simulations of Papers \citetalias{Fabbian+2010} and
\citetalias{Fabbian+2012},\\

\noindent (1) we attempt an absolute match of the synthetic continuum
radiation to disc-centre FTS observations;

\noindent (2) we carry out a study of the continuum intensity variation when
moving from disc centre to limb; and

\noindent (3) we investigate whether those models also correctly reproduce
selected [O\,\textsc{i}] absorption lines in the solar spectrum.

Concerning item (1), \citet{Trujillo-BuenoShchukina2009} used a single
3D HD snapshot by \citet{Asplund+2000} to produce a synthetic
disc-centre continuum intensity spectrum. Values at wavelengths in the
visible part of the spectrum proved to be roughly $10 \%$ below
observations.  In Papers \citetalias{Fabbian+2010} and
\citetalias{Fabbian+2012}, using 1.5D radiation transfer calculations,
we obtained a match of the absolute values of the synthetic continuum
intensity at disc centre to within $\sim 7 \%$ of the solar
observational data. In the present work, we see how this match can be
substantially improved by using a different spectral synthesis code
(\textsc{RH}, \citealt{Uitenbroek2001,Uitenbroek2002}), with,
therefore, opacities different to those used in our previous papers.

Concerning item (2), i.~e. the continuum CLV behaviour,
\citet{Koesterke+2008} found that shortcomings in CLV results from the
3D HD model of \citet{Asplund+2000} were roughly comparable to the
shortcomings of classical 1D models. The results indicated the
presence of too steep a temperature gradient in the 3D HD model. They
tentatively concluded that the improved description of the convective
energy transport in 3D may be offset by deficiencies introduced by a
too simplified radiation transfer scheme.
Including more opacity bins and additional sources of continuum
opacities led \citet{Pereira+2013a} (called Paper
\citetalias{Pereira+2013a} in the following) to achieve very good
agreement of their CLV results with observations. It is thus of
interest to check how well our 3D MHD models match the observations
concerning the continuum CLV behaviour, in spite of the reduced number
of opacity bins, by using the 3D radiation transfer capabilities of
the RH code.

In the final part of the current work, we will look at the effect of
the magnetic fields in convection models on the formation of selected
[O\,\textsc{i}] spectral lines. \citeauthor{Pereira+2013a} have
recently argued in Paper \citetalias{Pereira+2013a} that claims for
revised abundances based on 3D MHD modelling are premature.
On the other hand, \citet{Socas-Navarro2014} very recently found an
extreme sensitivity of the derived solar oxygen abundance to possible
systematics (retrieving values in the range $\log \epsilon
(\textrm{O})_{\odot} = 8.70-8.90$~dex). Typical uncertainties, mainly
intrinsic to employing atmospheric models, would thus be substantially
higher than what is usually claimed in the corresponding
literature. It is clear that the solar oxygen chemical abundance
derivation is particularly difficult, because the corresponding
absorption lines present in the solar spectrum are few and problematic
(see review by \citealt{Asplund2005}).

In this work, we will show that, barred the remaining uncertainties
(on, e.~g., continuum opacity, treatment of radiative transfer and
non-local thermodynamic equilibrium effects)
that complicate the treatment of spectrum formation, as well as those
affecting the solar spectrum observations \citep[see,
  e.~g.,][]{Beck+2011}, the MHD approach provides synthetic spectra
that match very well both the solar continuum intensity at different
inclination angles, as well as the [O\,\textsc{i}] $557$~nm and
$630$~nm absorption lines.

In the following section, we explain the simulation setup, we inform
on the input data chosen for the spectral synthesis computations, and
briefly present the method adopted in this study. In
sections~\ref{sec:abscontint}, \ref{sec:contintCLV} and
\ref{sec:spectral_lines}, we present the analysis of the theoretical
results of interest on disc--centre absolute continuum intensity, on
centre--to--limb variation of the continuum intensity, and on
continuum intensity normalised spectral line profiles, respectively,
and compare them to available observational data. The paper ends with
a discussion and conclusions (Sec.~\ref{sec:discussion}~and
\ref{sec:conclusions}).

\section{Simulation setup and resulting magnetic field
  stratification}\label{sec:setup_and_Bstratification}

\subsection{Setup of the numerical (M)HD simulations}\label{subsec:setup}

The 3D model atmospheres adopted as input for the spectral synthesis
are based on snapshots selected from the statistically stationary
regime of the convection in our previous (M)HD simulations (Papers
\citetalias{Fabbian+2010} and \citetalias{Fabbian+2012}), which used
input solar abundances of carbon, oxygen and iron of, respectively,
$\log \epsilon (\textrm{C})_{\odot}=8.55$, $\log \epsilon
(\textrm{O})_{\odot}=8.77$, and $\log \epsilon
(\textrm{Fe})_{\odot}=7.50$.  The snapshots we employ in this study
belong to the different MHD series in those papers, which had a
vertical magnetic field of uniform strength implanted at time $t=0$ of
$56$~G, $112$~G and $224$~G, respectively (labelled ``$50$~G'',
``$100$~G'' and ``$200$~G'' in the following).  Additionally, we will
also use here the ``$0$ G'', i.e., purely $HD$, series of those
papers.

The atomic and molecular transition data we employ in the intensity
spectrum calculations are from the \textit{Vienna Atomic Line
  Database} (\textit{VALD}) and from the \textit{National Institute of
  Standards and Technology Atomic Spectra Database} (\textit{NIST
  ASD}). VALD and its development have been described by
\citet{Piskunov+1995}, \citet{Ryabchikova+1999}, and
\citet{Kupka+1999}. In particular, we made use of the recent (February
2014) third release of
VALD\footnote{http://vald.astro.univie.ac.at/~vald3/php/vald.php}, and
of NIST ASD\footnote{http://physics.nist.gov/asd} in its online
version 5 \citep{Kramida+2013}.

We employed version 2 of the \textsc{RH} code
(\citealt{Uitenbroek2001,Uitenbroek2002}) to perform the 3D radiation
transfer calculations (we refer to them as ``R3D'' in the rest of this
article) on the basis of the input 3D (M)HD snapshots.
In the solar atmosphere, both Rayleigh scattering (off neutral
  hydrogen and helium) and Thomson scattering (off free electrons) may cause
  a departure from LTE conditions even in the absence of NLTE spectral
  lines. These scattering effects are negligible for the continuum at disc
  centre, but may become progressively more important towards the limb, where
  higher layers are being sampled. We thus include Rayleigh-Thomson (RT)
  scattering in the calculations of this paper, with a view to the
  centre-to-limb variation calculations of Sect.~\ref{sec:contintCLV}. 
  Vertically, we cut the input model to 64 depth points covering a
  physical depth of -0.5 to 0.5 Mm approximately. Horizontally, the
  full sampling of the original snapshots was maintained for the
  centre--to--limb variation (CLV) runs, with 252 grid points over 6
  Mm in each direction, while the sampling was reduced to 63 points in
  each horizontal direction over the same physical extent for the
  spectral line synthesis calculations.

We first computed the absolute intensity spectrum at disc--centre.
Then, we performed the CLV runs using a Gauss-Legendre set with 9
inclination angles and 4 azimuth angles per octant (thus
providing 36 angles per octant, or 144 rays in total).
In sections \ref{sec:abscontint} and \ref{sec:contintCLV} we will be
comparing the RH results of this paper with those of simpler (e.~g.,
1.5D, or without RT scattering) radiation transfer calculations
carried out with other codes: the corresponding details will be
explained in those sections.
For the spectral line calculations, which focused on the $557$~nm and the
$630$~nm oxygen features, the radiative transfer problem was solved
using a fine ($10$~m\AA) and uniform sampling in $\lambda$ space. The resulting output spectrum was set to cover 1.2 \AA\, around each of the two oxygen spectral lines of interest (which allowed us to account for any significant blends
affecting them).

The spectral synthesis results for the MHD case were obtained
neglecting the spectral line broadening due to the Zeeman effect, on
the basis of the expectation that, compared to the indirect effect due
to temperature changes, this direct effect of magnetic fields on
spectral lines is relatively small. It in fact
increases with $\lambda^{2}$ and it thus generally becomes appreciable
in the infrared (IR), rather than in the visible wavelength range
where the spectral lines we selected for this study fall.

\subsection{Average magnetic field distribution in the convection cells}
\label{subsec:Bstratification}

\begin{figure*}[ht]
\centering \epsscale{1.1}
\plotone{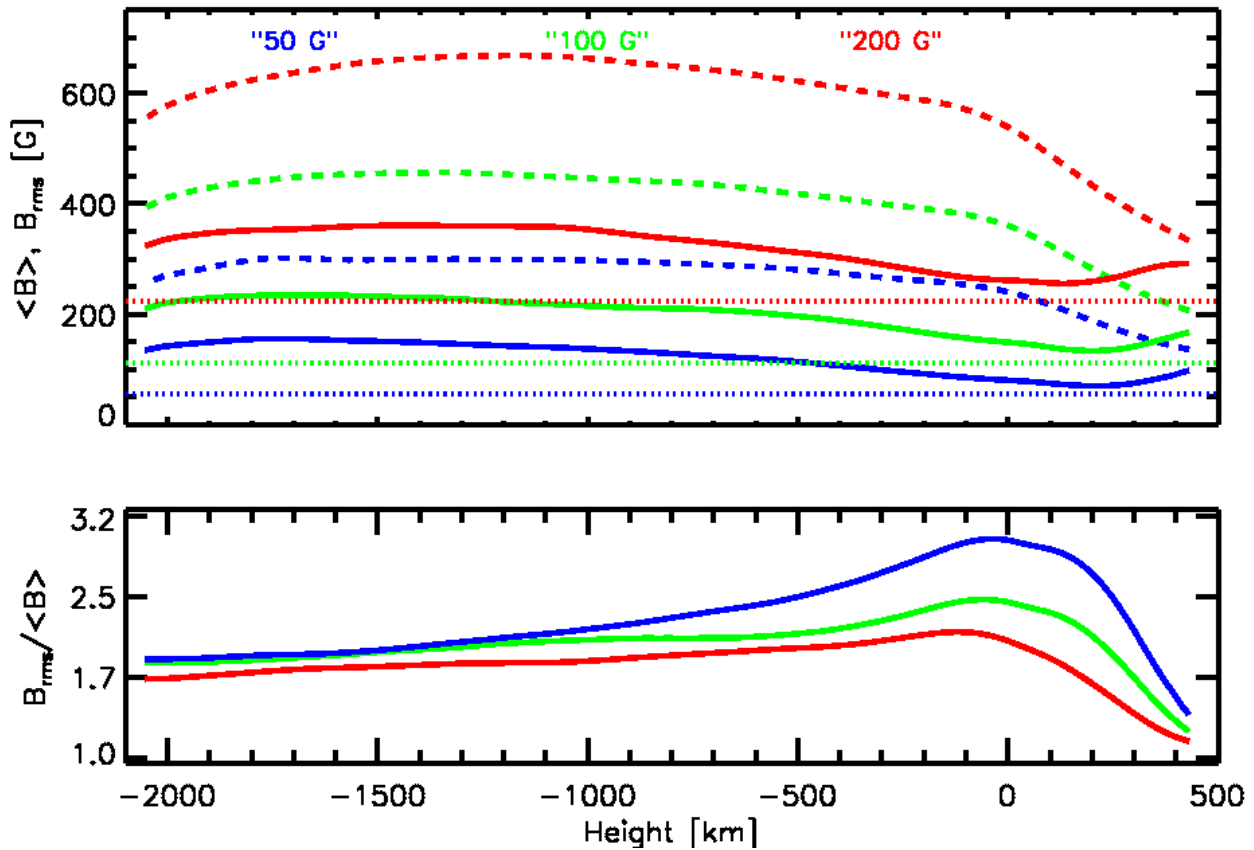}
\caption{Upper panel: height-stratification of the average magnetic
  field intensity (solid curve) and of the root-mean-square magnetic field 
  (thick dashed curve), expressed in Gauss units. Blue colour is used
  for the ``$50$~G'' series, green for the ``$100$~G'' series,
  and red for the ``$200$~G'' series. The
  dotted horizontal lines mark the constant value of the
  horizontally-averaged vertical component of the signed magnetic
  field intensity for the three series. Lower panel:
  height-stratification of the ratio of the rms magnetic field
  intensity, $\Brms$, and of the average magnetic field intensity
  $<B>$. Same colour coding as for the upper panel.
}
\label{fig01:Bfield}
\end{figure*}

Some of the results presented in 
  this paper are associated
  with the introduction of magnetic field in convection models, which can induce significant stratification changes, with consequent effects on spectral lines. Thus, for a better
  appreciation of our findings, it is adequate to consider
  briefly the average values of the magnetic field in the snapshots that
  serve as a basis for the radiation transfer calculations.  The upper panel
  of Fig.~\ref{fig01:Bfield} shows, using solid curves, the height
  distribution of the average magnetic field intensity, with the average
  being carried out separately for each horizontal plane and along time; the
  figure, therefore, shows the discrete version of:

\begin{equation}
\left<\, B \,\right>(z) 
= {1 \over t_f-t_0}{1\over A}\int_{t_0}^{t_f} dt \int^{}_{\rm \vtop{\hsize 1.3cm
\scriptsize
\vskip 0.mm\centerline{horizontal}\centerline{plane}}} \hskip -9mm
dx\,dy\; B(x,y,z,t) \;\,,  
\end{equation} 

\noindent with $A$ the area of the horizontal plane in the numerical box and
$t_0$ and $t_f$ the limits of the time interval used for the averaging.
\noindent Here, and for all subsequent figures in this subsection, the
time averaging is done using the last $100$ snapshots (corresponding
to $50$ solar minutes) of the statistically stationary regime in the
simulations, with snapshots separated by $30$ solar seconds. The
different series are identified with blue, green and
red colour for the ``$50$~G'', ``$100$~G'' and ``$200$~G'' series,
respectively. 
All three $\left< B \right>$ distributions decrease smoothly with
height in the topmost Megameter of the convection zone and
reach values of $\sim 80$~G, $\sim 150$~G, and $260$~G, respectively,
at $z=0$.
In the highest layers (i.e., roughly in the topmost $150$~km), the
distributions show an upturn, with the value of the mean magnetic field
strength increasing to a maximum of $\sim 100$~G, $\sim 170$~G, and $\sim
300$~G, respectively for the ``$50$~G'', ``$100$~G'' and ``$200$~G'' series.

It is worth noting that the ``$100$~G'' $\left< B\right>$-distribution 
of Fig.~\ref{fig01:Bfield} is not far from the distribution shown by
\citet{ShchukinaTrujillo-Bueno2011, ShchukinaTrujillo-Bueno2015} to match
the scattering polarization signal (induced by the Hanle effect) 
from a height of around $250-300$~km \citep{GurtovenkoKostik1989}, 
as derived from observations of the Sr\,\textsc{i} $460.7$~nm line, 
simultaneously allowing for the explanation of the polarization signal 
(induced by the Zeeman effect) from a height of around $60$~km, 
as derived from observations of the Fe\,\textsc{i} $630.25$~nm line. 
The small-scale local dynamo models of  
\citet{VoglerSchussler2007} or \citet{Rempel2014}, instead, lead to too weak 
average magnetic field strength values to explain the Sr\,\textsc{i} $460.7$~nm Hanle-effect data.

The upper panel of Fig.~\ref{fig01:Bfield} also shows (thick dashed
curves) the height dependence of the root mean square (rms) magnetic
field intensity, $\Brms = \sqrt{\left< B^{2} \right>}$,
which also serves to illustrate the distribution of magnetic energy in
the box in the stationary regime.  The $\Brms$ curves peak (with
values of $\sim 300$~G, $\sim 460$~G, and $\sim 670$~G, respectively)
in progressively higher layers (between $\sim 1700$ and $\sim 1200$
km) as magnetic flux increases between the ``$50$~G'' and the
``$200$~G'' case, then decrease monotonically. 

The dotted horizontal lines in the figure mark the constant value of the
vertical component of the field that was implanted at $t=0$ in each series.
In all cases, in the stationary regime the $\Brms$ field is larger than the
average field intensity. This is a result of the intermittency created in the
field by the action of the convective flows. To show this effect from a
different perspective, we plot in the lower panel of Fig.~\ref{fig01:Bfield}
the ratio $\sqrt{\left< B^{2} \right>} / \left< B \right>$.
This ratio can serve as an index for the intermittency level in each
of the time-series.  The ratio is seen to increase monotonically with
height in the subsurface layers, and to reach a maximum very near the
surface, $z=0$. This is expected since the Alfven Mach number of the
convection flows is largest near the surface. We also see that the
intermittency level decreases the larger the initially implanted
field, with peak ratios of $\sim 3.0$, $2.4$ and $2.1$ for the
``$50$~G'', ``$100$~G'' and ``$200$~G'' series, respectively. Above
the surface, the intermittency ratio decreases sharply, given the less
passive role of the magnetic field generally for higher atmospheric
levels and the decreasing intensity of the convection there.

\section{Disc--centre absolute continuum intensity}\label{sec:abscontint}

\begin{figure*}[ht]
\centering
\epsscale{1.1}
\plotone{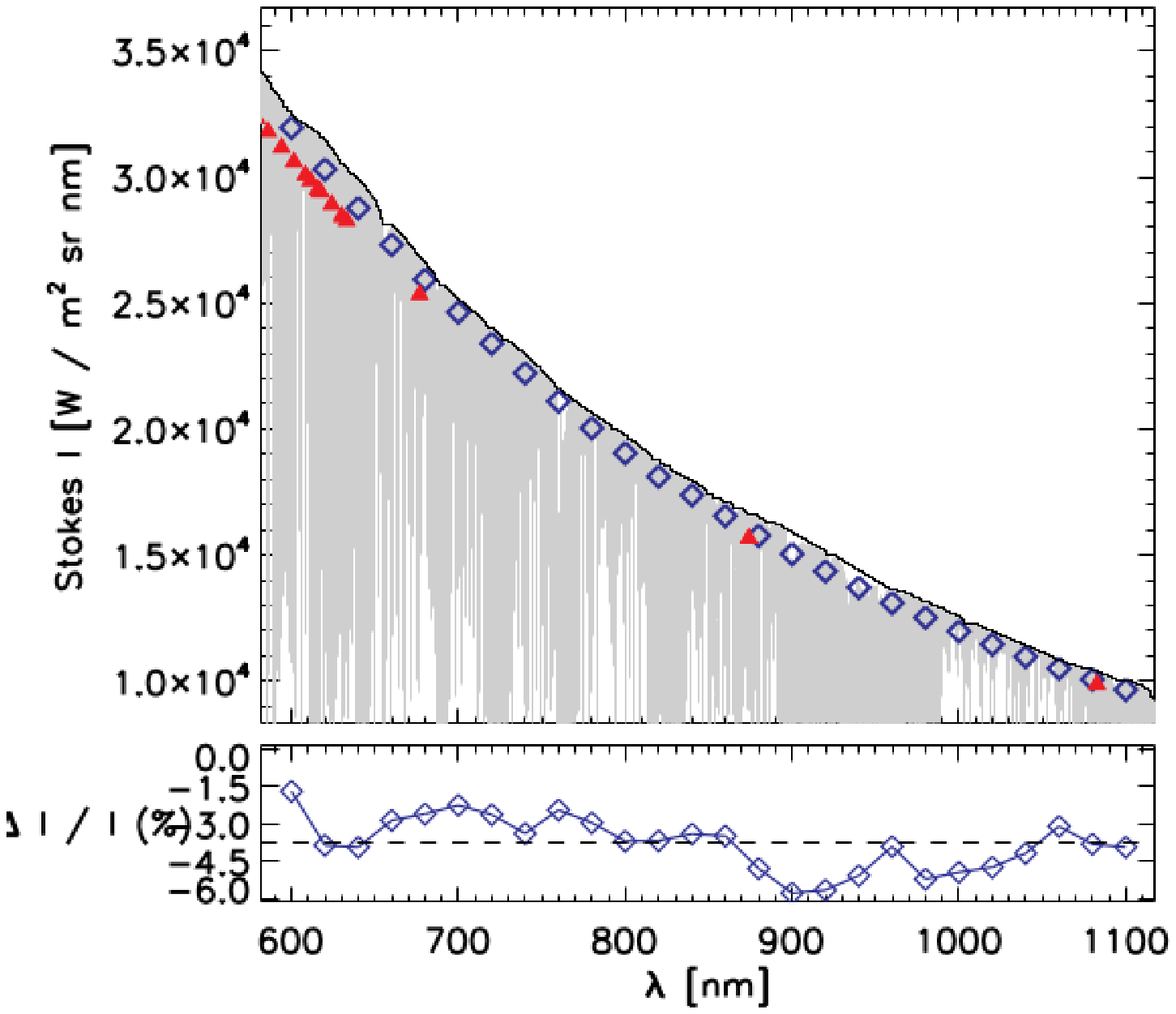}
\caption{\textit{Absolute} disc--centre continuum intensity at
  different wavelengths (upper panel). The solar atlas data from FTS
  observations are shown in light grey colour. The theoretical R3D
  values are plotted as blue diamond symbols. Our FTS pseudocontinuum
  is shown as a solid black curve. Our R1.5D values of Paper
  \citetalias{Fabbian+2012} are shown as red triangles. In the lower
  panel, we show the percentage difference between the R3D results and
  the FTS pseudocontinuum. The dashed horizontal line marks the $\sim
  -3.8\%$ average (R3D MHD - FTS) deviation level.
}
\label{fig1:simsVSobs_abscontint_disc-centre}
\end{figure*}

In this section we present the first group of results in the paper,
namely the comparison with observations of the \textit{absolute}
disc--centre continuum intensity values obtained through spectral
synthesis calculations from our (M)HD snapshots
(Fig.~\ref{fig1:simsVSobs_abscontint_disc-centre}, upper panel). For
the observations we took the \textit{Fourier Transform Spectrograph}
(FTS) data, drawn in the figure as a collection of points in light
gray. For the comparison, a pseudocontinuum is needed, to use as an
envelope to the FTS data. We obtained it (black line) by linking the
local maxima of the original FTS data, bridging gaps across wavelength
regions affected by strong absorption features. Our R3D data are shown
as blue diamonds and they represent the absolute continuum intensity
values obtained using the three-dimensional radiation transfer
calculation explained in Subsection \ref{subsec:setup}. To take into
account the fluctuations in time, we calculated the value of the
continuum at each of the wavelengths as the time average of 21
snapshots covering $50$ solar minutes, with the first snapshot taken
at $100$ solar minutes inside the statistically stationary part of the
simulation. The theoretical values (the blue diamonds) are
very near the FTS continuum, as also apparent in the lower panel of the
figure: the average mutual deviation is $\sim 3.8\%$, with the minimum
and maximum relative difference being $\sim 1.5\%$ (at $600$ nm) and
$\sim 6\%$ (at $900$ nm), respectively.

In the upper panel of Fig.~\ref{fig1:simsVSobs_abscontint_disc-centre}, we
also show, as red triangles, our previous results of Paper
  \citetalias{Fabbian+2012}, obtained with a different radiation transfer
  calculation, namely, using the Lilia code of \citet{Socas-Navarro2001} and
  without any RT scattering.  The better match to observations of our current
  results compared with those of Paper \citetalias{Fabbian+2012} provides
  reassurance both on the quality of our models and on the opacities used by
  the \textsc{RH} code. We further tested the effect of not including
  Rayleigh and Thomson scattering in our current RH calculations.  We found
  the change in absolute continuum intensity to be unimportant. 

  The small remaining percentage difference (shown in
  Fig.~\ref{fig1:simsVSobs_abscontint_disc-centre}, lower panel) of
  the R3D results with respect to FTS can be understood as partly due
  to the fact that the (M)HD snapshots have a slightly lower (within
  $\lesssim 20$\,K) outgoing radiation flux compared to the measured
  solar one.  However, that can explain only $\lesssim 1\%$ of the
  discrepancy.  A more significant contribution to the $\sim 3.8\%$
  average difference with FTS may arise from the comparatively low
  number of opacity bins used in our Stagger-code calculation. A
  thorough treatment of line blocking/blanketing effects by using very
  accurate atomic data for many spectral lines may also provide a
  further improvement of the match with observations, an approach
  which can, however, become prohibitively time-consuming when coupled
  with 3D radiation transfer, due to the plethora of lines present in
  the solar spectrum that need to be taken into account. In
    any case, the FTS atlas, like any observation, is, itself,
    affected by uncertainties of up to a few
      percent (for example, affecting regions of strong telluric
      absorption, see \citet{Burlov-VasiljevMatvejevVasiljeva1998}),
      which contribute to part of the difference seen here.

The comparison between the theoretical results for HD and MHD runs is
also of interest. It turns out that there is only a small difference
between them; in fact, the HD curve would be nearly indistinguishable
if plotted alongside the blue diamonds of
Fig.~\ref{fig1:simsVSobs_abscontint_disc-centre} (which correspond to the
R3D MHD results for the ``$200$~G'' case): the HD continuum is only $\Delta(I)
\lesssim 0.02 
\cdot 10^{4}$ W m$^{-2}$ sr nm below the MHD one. The reason why their
continuum level turns out to be nearly identical is that (Paper
\citetalias[][Fig.~3]{Fabbian+2010}; Paper
\citetalias[][Fig.~1]{Fabbian+2012}) the MHD-HD temperature difference
tends to be smallest (tending to negligible) precisely in the layers
where the continuum forms.

\section{Centre--to--limb variation of the continuum
  intensity}\label{sec:contintCLV}

As an observer's line-of-sight moves towards the solar limb, the
continuum intensity decreases because it originates in higher layers
which are not as bright (warm) as the deeper ones to which the
observer's line-of-sight penetrates when looking at disc centre
\citep[e.g.,][chapter 9]{Gray1992}. We can thus use the behaviour of
the theoretical CLV to verify whether the temperature stratification
of the models results in a proper matching of the solar continuum
intensity CLV.

\begin{figure*}[ht]
\centering
\epsscale{1.0}
\plotone{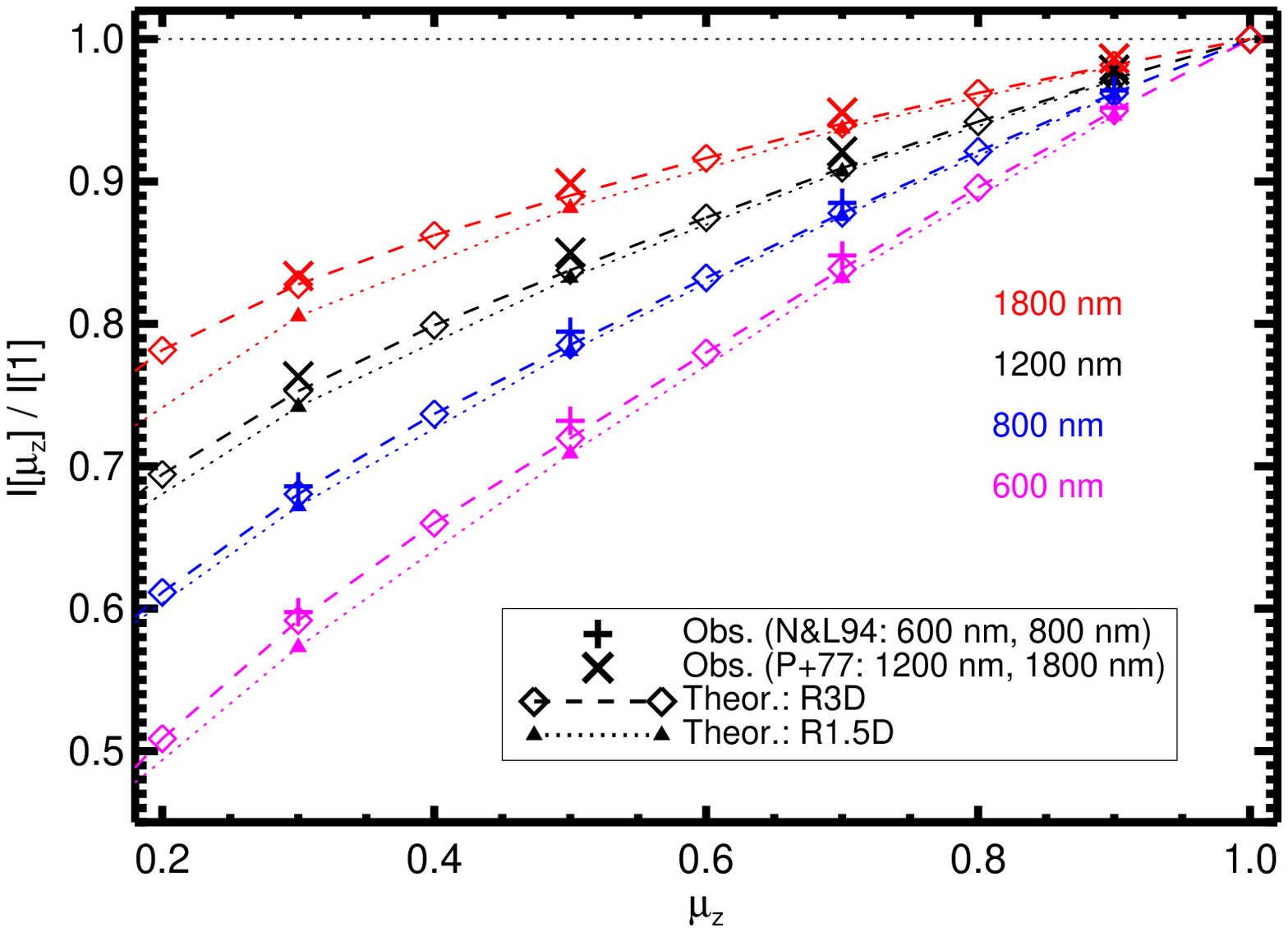}
\plotone{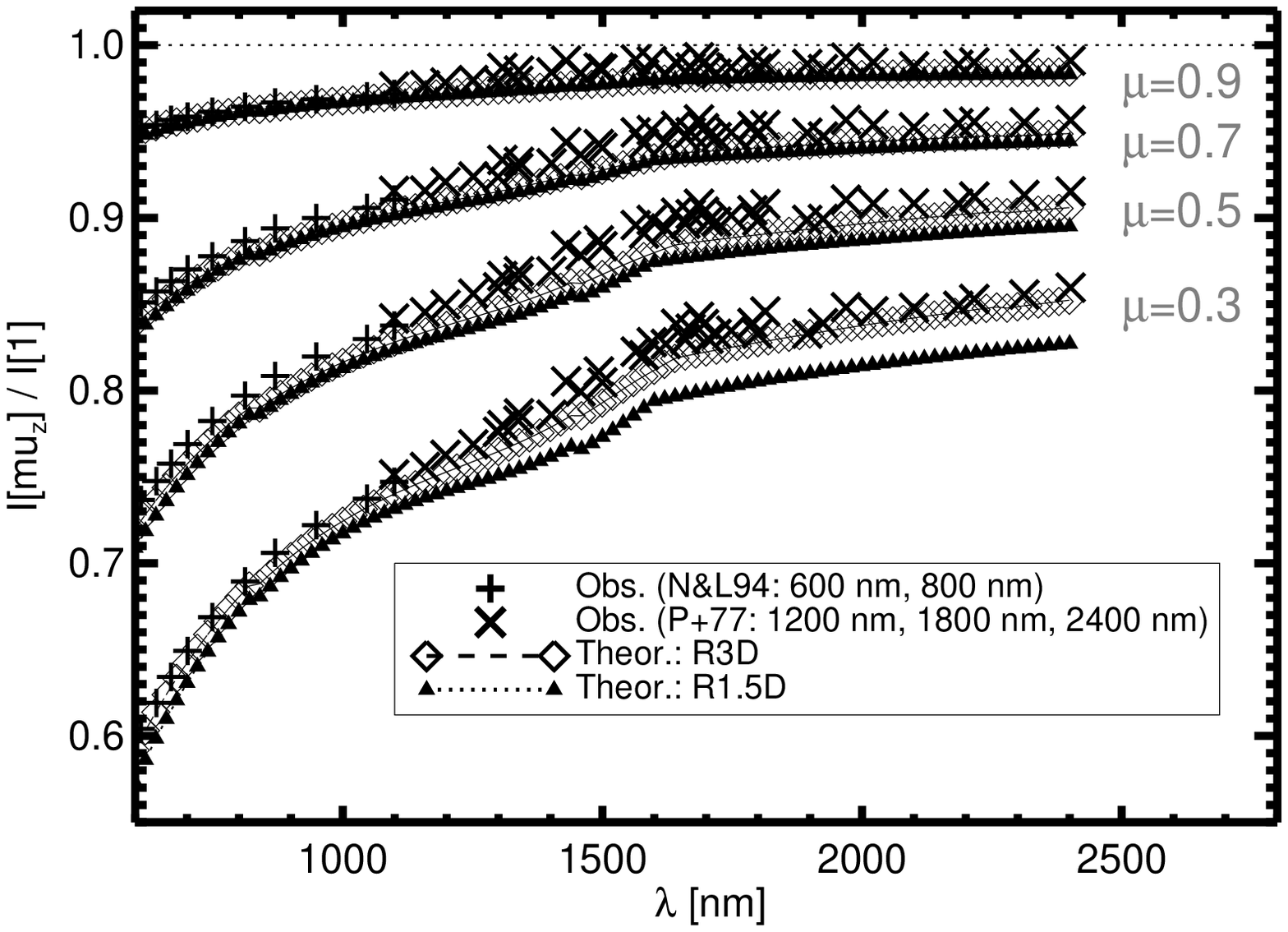}
\caption{\textit{Upper panel}: predicted continuum intensity at
  different wavelengths (indicated by the different colours, as
  labelled) from our R3D (this paper, diamond symbols linked by 
  dashed segments) and R1.5D \citepalias[Paper][triangular
  symbols]{Fabbian+2012} computations for different values of
  $\mu_{\textrm{z}}=\cos\theta$ (where $\theta$ is the heliocentric
  angle) is compared to solar limb-darkening observational data by
  \citet{NeckelLabs1994} ('+' symbols), and by
  \citet{PierceSlaughterWeinberger1977} ('$\times$' signs). The dotted
  horizontal line indicates the normalisation level with disc--centre
  continuum intensity. \textit{Lower panel:} the same, now versus
  wavelength and with the different curves being for different
  $\mu_{\textrm{z}}=\cos\theta$ values, as labelled. Symbols are the
  same as for the upper panel, as marked in the legend.
      }
  \label{fig2:simsVSobs_contint_CLV}
\end{figure*} 

In Fig.~\ref{fig2:simsVSobs_contint_CLV}, we show our results on
continuum intensity centre--to--limb variation (CLV) for the synthetic
spectra normalised to the corresponding value at disc centre. We
studied the continuum intensity behaviour both against
$\mu_{\textrm{z}}=\cos \theta_{\textrm{helio}}$ (upper panel), with
$\theta_{\textrm{helio}}$ the heliocentric angle, as well as against
wavelength (lower panel) in the range $600$~nm to $2400$~nm. In the
upper panel of Fig.~\ref{fig2:simsVSobs_contint_CLV}, the theoretical
R3D continuum intensity results from our computations for the MHD case
are represented as diamond symbols linked by dashed segments, while
observations\footnote{Note that the \citet{NeckelLabs1994} data agree
  well with the less recent ones by \citet{PierceSlaughter1977} (not
  shown here) in the wavelength interval of $303-730$~nm common to the
  two datasets. In the adjacent common wavelength interval of
  $740-1100$~nm, the \citet{NeckelLabs1994} data agree well with those
  of \citet{PierceSlaughterWeinberger1977}. For this reason, we show
  here the values for the latter dataset only for the non-overlapping
  wavelength region above $1100$~nm.} are shown as '+' signs for
\citet{NeckelLabs1994} and '$\times$' symbols, for
\citet{PierceSlaughterWeinberger1977}. Our R3D results are seen to
match very well the solar limb darkening observational data.
Taking into account radiation transfer effects properly in a 3D inhomogeneous 
atmosphere thus produces realistic values of the average radiation 
in the continuum.
In the figure we are also showing (as triangles linked by a dotted line) the
results obtained using a simplified radiation transfer 
calculation with the 
Nicole code of \citet{Socas-Navarro+2014} (which, like the Lilia code before,
uses a different opacity package than RH). The Nicole version used had a 1.5D
approach to the horizontal radiation transfer (called ``R1.5D'' in the
following). In that approach, we considered as many $\mu$ angles as in
the full 3D calculation just shown, but the horizontal radiation transfer
coming out of a given pixel is calculated via an infinite repetition of the
corresponding column in the two horizontal directions.  The R1.5D results 
are found to deviate substantially from the observational data: differently
from the previous Section's result on disc-centre continuum intensity, the
continuum CLV values are not only sensitive to adopted opacities, but are
also sensitive to horizontal transfer effects, especially as one moves toward
the limb. It is then natural that the R1.5D results deviate
  importantly from the observations when approaching the limb, since they 
do not take proper account of the
hills and valleys of the $\tau$ isosurfaces in the 3D models.

The lower panel of Fig.~\ref{fig2:simsVSobs_contint_CLV} compares the
observational data and theoretical results just discussed, now against
wavelength. This more clearly highlights the differences between the R3D and
R1.5D approaches. The symbols are
the same as in the upper panel of this figure (diamonds for R3D,
triangles for R1.5D, '+' signs and '$\times$' for the observations), 
except without colour coding in
wavelength. Again here, the R3D-R1.5D deviations (i.~e., those due to
horizontal transfer effects) generally increase when moving toward the
limb, in particular, at longer wavelengths. This is in line with the
results in Paper \citetalias{Pereira+2013a} (their Fig.~3), where a
similar behaviour is visible in the case of the classical MARCS and
PHOENIX 1D solar photosphere models (while the 1D Holweger \& M\"uller
model, being semiempirical, not surprisingly matches the observational
data as well as their 3D models). It should be noted that the
comparison of theoretical and observational data very close to the
limb (i.e., at $\mu_{z}<0.2$ or, equivalently,
$\theta_{\textrm{helio}} \gtrsim 80$~degrees) is less helpful, since
the latter data tend to become less reliable there.

In both panels of Fig.~\ref{fig2:simsVSobs_contint_CLV}, we show only the
``$200$~G'' MHD case, for both the R1.5D and R3D radiative transfer
calculations. The R3D case for the other series, in fact, 
generally displays a very similar match to observations as the one plotted. 
The situation is thus similar to what was already seen in the previous Section regarding
the absolute continuum intensity, with relatively small differences between
the results for the HD case and the MHD case(s).

The fine match to observations of our current R3D results again
provides reassurance on the accuracy of our 3D (M)HD models and on the
\textsc{RH} opacities and spectral synthesis. Differently from
Sec.~\ref{sec:abscontint}, though, it is harder to disentagle for
these CLV runs how much of the difference between R3D and R1.5D
results is due to, respectively, different opacities, Rayleigh and
Thomson scattering (which should play more of a role at inclined
angles), and pure geometrical / horizontal transfer (3D vs. 1.5D)
effects in the treatment of radiation.

\section{[O\,\textsc{i}] spectral line profiles}\label{sec:spectral_lines}

\begin{figure*}[ht]
    \begin{center} 
    \epsscale{1.0}
      \plotone{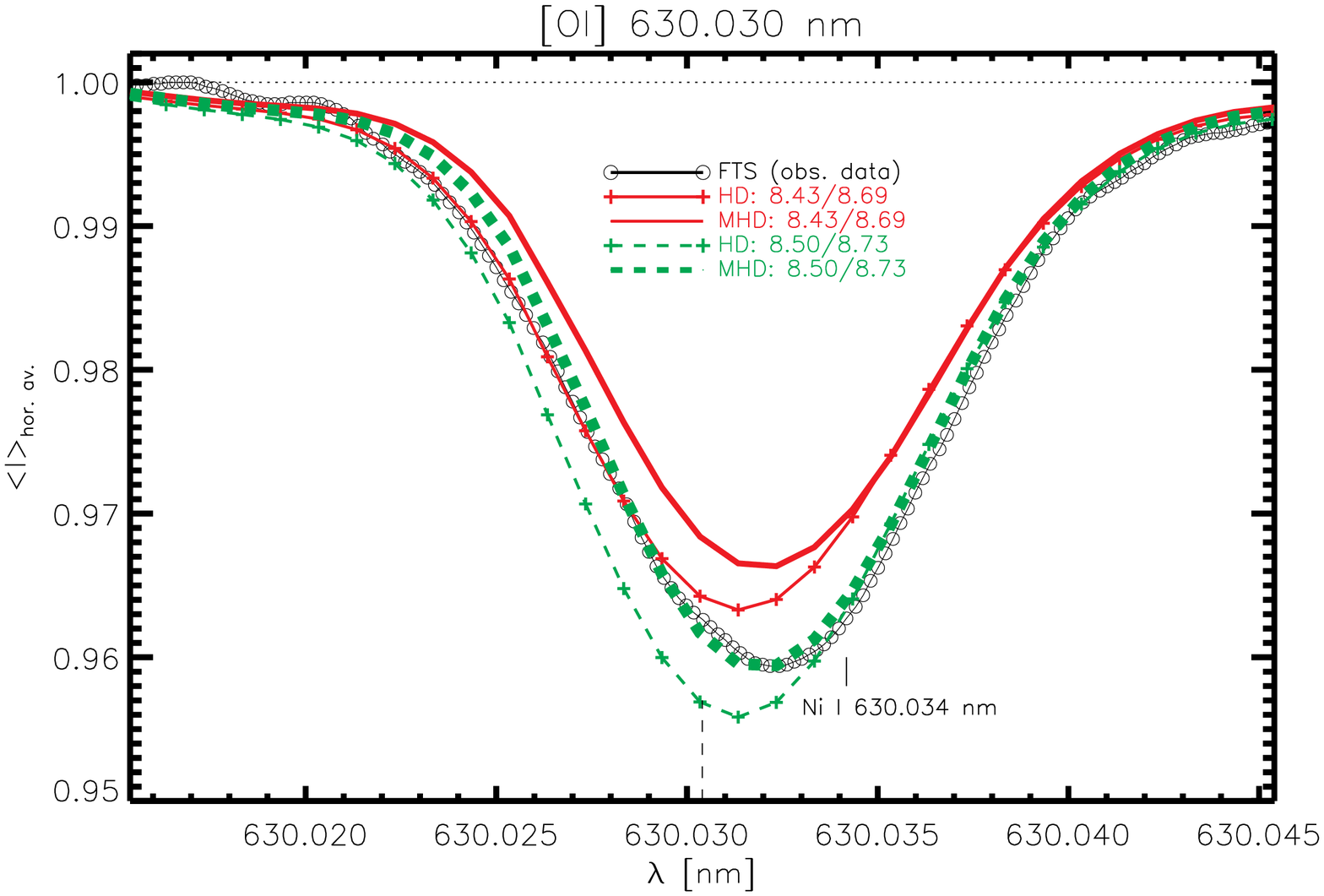}
    \end{center} 
    \caption{Solar observational disc--centre intensity data from FTS
      (black open circles) and R3D synthetic intensity profiles
      (coloured curves) in HD and MHD for the blended feature at
      $630$~nm, which contains the [O\,\textsc{i}] line of interest.
      As per the legend in the figure, the HD and MHD synthetic
      profiles shown are the result of adopting different choices for
      the input C and O solar abundances. The wavelength of the
      [O\,\textsc{i}] line and of the blending Ni {\sc i} line are
      marked with a dashed vertical segment and with a solid vertical
      segment, respectively. The dotted horizontal line indicates the
      normalisation level with the adjacent continuum intensity.}
    \label{fig3:simsVSobs_[OI]_630nm_disc-centre}
\end{figure*} 

\begin{figure*}[ht]
    \begin{center} 
    \epsscale{1.0}
      \plotone{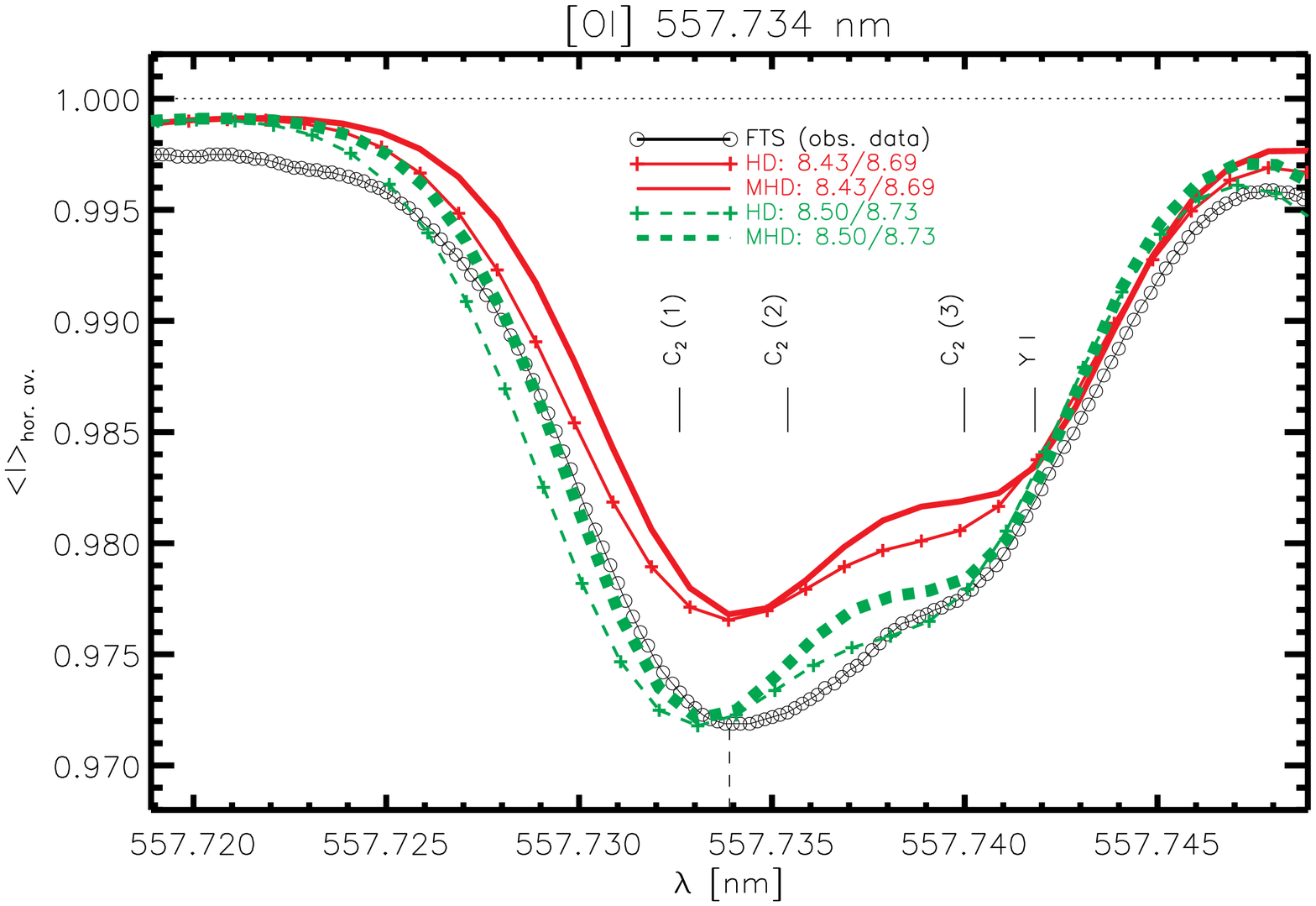}
    \end{center} 
    \caption{Solar observational disc--centre intensity data from FTS
      (black open circles) and R3D synthetic intensity profiles
      (coloured curves) in HD and MHD for the blended feature at
      $557$~nm, which contains the [O\,\textsc{i}] line of interest.
      As per the legend in the figure, the HD and MHD synthetic
      profiles shown are the result of adopting different choices for
      the input C and O solar abundances. The wavelengths of the
      [O\,\textsc{i}] line is marked with a dashed vertical segment,
      and the wavelengths of important blending spectral lines are
      marked with solid vertical segments ($\textrm{C}_{2} (1) =
      \textrm{C}_{2}\ 557.733$~nm; $\textrm{C}_{2}\, (2) =
      \textrm{C}_{2}\ 557.735$~nm; $\textrm{C}_{2}\, (3) =
      \textrm{C}_{2}\ 557.740$~nm; $\textrm{Y\,\textsc{i}} =
      \textrm{Y\,\textsc{i}}\ 557.742$~nm). The dotted horizontal line
      indicates the normalisation level with the adjacent continuum
      intensity.}
    \label{fig4:simsVSobs_[OI]_557nm_disc-centre}
\end{figure*}

In this section, we analyse the behaviour of the profile of selected
spectral absorption lines when using 3D snapshots from our HD and MHD
series and when changing the C and O solar abundance adopted as input
to the spectral synthesis code. The main aim is to understand how the
change in temperature stratification due to the presence of magnetic
fields affects the spectral lines and to additionally verify whether
our 3D MHD photospheric models coupled with 3D radiative transfer for
the \textit{a posteriori} spectral synthesis can fit observations
better, or at least as well as when using input photospheric models
from our HD series. For this part of our study, we selected two
representative spectral lines of neutral atomic oxygen, namely
[O\,\textsc{i}] $557$\, and [O\,\textsc{i}] $630$\,nm, both produced
by radiatively forbidden transitions. These two spectral lines are
deep--forming and very weak in the solar spectrum.  According to
\citet{GurtovenkoKostik1989}, the main contribution to the formation
of their core is from a range of layers around a height of,
respectively, $\sim 130$\,km and $\sim 180$\,km, above
$\tau_{500\,\textrm{nm}}=1$. Being formed in LTE, and given the
scarcity of available oxygen lines in the solar spectrum, these
spectral lines represent important tools in terms of the solar oxygen
abundance determination \citep{MelendezAsplund2008,Scott+2009}.  Their
use for the purpose of estimating the solar oxygen abundance is,
however, significantly affected by blends, mainly C$_{2}$ molecular
lines for the [O\,\textsc{i}] $557$\,nm absorption line, and a
Ni\,\textsc{i} atomic line for [O\,\textsc{i}] $630$\,nm.
Additionally, the blending Ni\,\textsc{i} line requires proper
treatment of its isotopic splitting, accounting for which
significantly improves the fit to observational data. In the
  following analysis, we include all significant blends, among them the
  C$_{2}$ and Ni\,\textsc{i} lines just mentioned.
Note that, while
free of significant known blends, the strong, widely used neutral
oxygen triplet lines at $777$\,nm are prone to significant non--LTE
effects \citep[see][]{Fabbian+2009,Pereira+2009} and thus would
require a more complicated and time-consuming line synthesis treatment
than the one we adopt here.

The [O\,\textsc{i}] $557$\, and [O\,\textsc{i}] $630$\,nm lines fall
in the wavelength region where the \textit{absolute} continuum
intensity from our theoretical R3D calculations best matches the FTS
solar atlas (see Fig.~\ref{fig1:simsVSobs_abscontint_disc-centre}). We
here compare the \textit{continuum--normalised} theoretical spectra
with \textit{Fourier Transform Spectrograph} (FTS) observations of the
solar spectrum. We normalised the spectral intensity across each of
the wavelength regions to the relevant intensity of the adjacent
pseudocontinuum. Note that for spectral lines around $600$~nm, the
variation of the continuum level with wavelength across the
line is negligible ($<10^{-4}$) and does not need to be taken into
account for our purposes.

The line formation calculations are computationally demanding,
  since many wavelength points must be used (as explained in
  Sec.~\ref{subsec:setup}) for an accurate study of the line profiles. 
Therefore, we selected as input model photosphere a
  single snapshot per series, namely one from the series with no magnetic
field and another one from the ``$200$~G'' MHD series, respectively labelled
``HD'' and ``MHD'' in the following. Both snapshots were taken
at a point in the calculations which corresponds to $100$ minutes of solar
time evolution in the statistically stationary regime of the convection. The
choice to study the relevant effects based on the ``$200$~G'' MHD series was
made to estimate the expected \textit{maximum} changes that the presence of
magnetic fields induces -- via a modified atmospheric temperature structure
-- on the profile of the selected oxygen spectral lines.  Given the weakness
of the latter, this also makes the representation of results in the figures
of this Section clearer. This study thus covers quiet Sun to weak plage
conditions.

In Figs.~\ref{fig3:simsVSobs_[OI]_630nm_disc-centre} and
~\ref{fig4:simsVSobs_[OI]_557nm_disc-centre} we show the solar
disc--centre intensity results of the R3D line synthesis for,
respectively, the spectral region around the [O\,\textsc{i}] $630$~nm
and $557$~nm absorption features. Our initial approach was to attempt
to fit the FTS observations with the R3D synthetic spectra obtained
using input abundance values from \citet{Asplund+2009}, in particular,
$\log \epsilon (\textrm{C})_{\odot}=8.43$~dex and $\log \epsilon
(\textrm{O})_{\odot}=8.69$~dex. The solar carbon abundance
  adopted is important here, since it affects the right calculation of
  the C$_{2}$ concentration, which in turn is relevant due to the
  corresponding molecular blends affecting the [O\,\textsc{i}]
  $557$\,nm spectral feature. The formation of CO, instead, 
    takes place at cooler temperatures than relevant here and is
    therefore unimportant for our results.

By comparing the corresponding HD and MHD synthetic intensity profiles
(red curves in Figs.~\ref{fig3:simsVSobs_[OI]_630nm_disc-centre} and
~\ref{fig4:simsVSobs_[OI]_557nm_disc-centre}, respectively for
[O\,\textsc{i}] $630$\,nm, and [O\,\textsc{i}] $557$\,nm), one clearly
sees that both of the spectral absorption features we targetted become
\textit{weaker} in MHD, as expected due to the higher average
temperature in the relevant line-forming layers. On the other hand,
for both spectral lines, the fit against observations is not
satisfactory when using those abundances.  The lines are significantly
too weak, in both MHD and HD.  Not including blends would of course
produce spectral profiles having the wrong shape because they would
miss the additional asymmetry.

The poor fit of the red curves in
Figs.~\ref{fig3:simsVSobs_[OI]_630nm_disc-centre} and
~\ref{fig4:simsVSobs_[OI]_557nm_disc-centre} is to be expected because
our (M)HD models have a different temperature stratification from the
ones employed by \citet{Asplund+2009}. One may thus change the value
of the elemental abundances adopted as input in the spectral
synthesis, to verify the sensitivity of spectral features to a
different chemical composition.
We tested adopting different C and O input solar abundances and
visually obtained a best fit when choosing the values $\log \epsilon
(\textrm{C})_{\odot}=8.50$~dex and $\log \epsilon
(\textrm{O})_{\odot}=8.73$~dex. The green dashed curves in
Fig.~\ref{fig3:simsVSobs_[OI]_630nm_disc-centre} and
Fig.~\ref{fig4:simsVSobs_[OI]_557nm_disc-centre} show the profile of
the targetted spectral lines when adopting these as input abundance
values. The thick dashed curve represents the MHD case, while the thin
curve with crosses is for the HD case. As seen in the figures, for
both of the wavelength regions studied here, our synthetic lines for
the MHD case match very well the disc--centre intensity profiles
observed from the Sun, with the fit being particularly good in the
case of the $630$~nm spectral absorption feature. It is noteworthy
that this good agreement in both wavelength regions is achieved while
using the same set of C and O solar abundance values, as one should
require for consistency. Note that the abundance of Ni was maintained
fixed at the value of \citet{Asplund+2009}, $\log \epsilon
(\textrm{Ni})_{\odot}=6.22$~dex.

In the case of the $630$~nm absorption feature, a crucial contribution
to the composite intensity profile comes from accounting for the
isotopic splitting of a blending neutral nickel feature present on its
longer-wavelength side.
We achieved this by scaling the Johansson et al. (2003) measured $\log
gf$ for the nickel spectral line under consideration by the abundance
of each of five Ni\,\textsc{i} isotopes (the same treatment adopted by
\citealt{Pereira+2009}, see their Sect. 4.2.2 and their Table 4).
Regarding the $557$~nm absorption feature, the presence of a small
residual discrepancy -- with respect to the FTS data -- on the red
side (at $\sim 557.738$~nm) of our R3D line profile obtained using our
final adopted abundances, and seen in
Fig.~\ref{fig4:simsVSobs_[OI]_557nm_disc-centre}, may indicate that a
weak, yet unknown blend might be present at that wavelength, just
redward of the [O\,\textsc{i}] line.  We note that
\citet{MelendezAsplund2008} obtained a very good fit of the $557$~nm
wavelength region, at least in the case of the Delbouille solar atlas,
when using their own spectrum synthesis code on a three-dimensional
hydrodynamical model of the solar atmosphere obtained by temporally
and spatially averaging the snapshots from the series of
\citet{Asplund+2004}.  They derived values of $\log \epsilon_{\textrm
  C} \sim 8.43$ based on fitting the blending C$_{2}$ lines in the
relevant wavelength region.  With that value, they then found $\log
\epsilon_{\textrm O} \sim 8.70$ from the $557$~nm absorption feature.
Those abundance values are significantly ($\sim 0.03-0.07$~dex) lower
than our final adopted input C and O solar abundances.

\ 

\section{Discussion}\label{sec:discussion} 

In this study, we have obtained a good match of quantities resulting from
our 3D (M)HD simulations against observational data.  The studied
quantities were: the \textit{absolute} continuum intensity at disc
centre, the CLV of the continuum intensity, and the profiles for two
selected absorption features, each containing an oyxgen spectral line
([O {\sc i}] $557$~nm and $630$~nm, respectively) blended with lines
of other chemical elements.

The continuum intensity (disc-centre and/or CLV) proves to be sensitive to
adopted opacities and 3D radiation transfer capabilities of the spectral
synthesis procedure, and corresponding observations at inclinations from
disc centre to close to the limb are very well fit by our 3D models
(Secs.~\ref{sec:abscontint} and \ref{sec:contintCLV}). The
good matching of absolute continuum intensity at disc centre
(Sec.~\ref{sec:abscontint}) confirms
that the effective temperature of the deep--lying continuum--forming
layers is realistic, but does not provide tight constraints for higher
layers. The CLV of continuum intensity is a better tool for that purpose,
and our results provide a
quite satisfactory match (Fig~\ref{fig2:simsVSobs_contint_CLV}) of
solar limb darkening data, with the implication that the theoretical
average temperature stratification should be relatively
close to the correct one. The synthetic CLV values tend to be
lower than the observations, in particular at intermediate angles
($\mu_{\textrm{z}} \sim 0.5$), when considering intensities normalized to the
values at disc centre, as is customary in CLV studies. 
So, this deviation adds up to the error in the absolute 
matching found in Sec.~\ref{sec:abscontint} (average of $4\%$, see
Fig.~\ref{fig1:simsVSobs_abscontint_disc-centre}). 
For an even more accurate fit of solar continuum data, lower
opacities and/or increased radiative flux at $\sim 1400-1800$~nm
(possibly due to using a larger number of opacity bins in the
simulations, or to atomic data improvements at those wavelengths), a
mildly warmer temperature structure especially in intermediate to deep
layers of the atmospheric model (thus, with a steeper temperature
gradient at $\log \tau_{500~nm} \gtrsim -1$) and/or more magnetic
flux, may be required.  Note that precisely in the wavelength range in
question the line-of-sight penetration in the Sun reaches its maximum
\citep[][chapter 9]{Gray1992}, which would indicate the need for
better modelling in those deep, warm layers.  Alternatively, the
larger discrepancy present especially around $1500$~nm (see
Fig.~\ref{fig2:simsVSobs_contint_CLV}, lower panel) may imply that
improved observations are required in that wavelength region.

Regarding the [O {\sc i}] spectral lines targetted in this study, we
obtain a good match of the synthetic data with the corresponding
observations. In particular, when employing the 3D MHD ``$200$~G''
photospheric model, the match is achieved for $\log
\epsilon_{\textrm{MHD}} (\textrm{C})_{\odot}=8.50$~dex and $\log
\epsilon_{\textrm{MHD}} (\textrm{O})_{\odot}=8.73$~dex, while neither
the HD nor the MHD models are able to fit observations if adopting the
lower solar abundances of \citet{Asplund+2009}, $\log \epsilon
(\textrm{C})_{\odot}=8.43$~dex and $\log \epsilon
(\textrm{O})_{\odot}=8.69$~dex. To appreciate the effect of the
magnetic fields with the largest possible clarity, in
Sec.~\ref{sec:spectral_lines} we only showed results for the extreme
cases of zero magnetic field and of the ``$200$~G'' MHD time-series,
the latter being the case with the highest vertical magnetic flux
among our ensemble of simulations.  Comparing these two cases, there
is a clear effect of the modified average atmospheric temperature
stratification on the line profile.  However, the other MHD series
(``$50$~G'' and ``$100$~G'') are more likely to be representative of
standard quiet Sun conditions. Additionally, for maximum accuracy when
aiming at absolute solar chemical composition determinations, line
broadening by the Zeeman effect should be included. Consequently, the
final solar carbon and oxygen abundances that we could deduce from the
present study may be somewhat different than obtained in 
Sec.~\ref{sec:spectral_lines}.
This needs to be checked in detail using a larger number of spectral
lines -- and, ideally, even better observational data -- for the chemical 
elements in question, in future studies. 
In any case, it
is a significant result that, due to the sensitivity of the profiles
to the temperature stratification, \textit{differential} effects are
present for the two oxygen lines studied, and that with an appropriate
choice of input chemical abundances, our MHD models can provide a
good match of the corresponding
observed spectra. Magnetic effects are therefore non-negligible, and given
the observational evidence of the presence of magnetic fields all over the
Sun, they should be taken into consideration. One may achieve a
match of the HD model to the observed line spectra by raising the
abundances a little from the level of \citet{Asplund+2009}, 
not so much as needed to achieve a good fit when using
an MHD case. For each particular case (HD or MHD) one
  should test if one could find a comparable level of consistency
  in the set of derived C and O solar abundance values as we have found here
  for the ``$200$~G'' MHD case. 

The fair match of different observational quantities is an important result
both in terms of confirming the quality of the MHD models we employ, as well
as in relation to the conclusion presented in Paper
\citetalias{Pereira+2013a} where its authors argued that it is still
premature to consider a revision of solar abundances based on 3D MHD
modelling.  Interestingly, although our 3D magnetoconvection calculations
employ the same code as used by them, our resulting photospheric models have
a different temperature stratification than theirs.  Differences in the
number of opacity bins used, equation-of-state treatment, spatial resolution
and temporal variation of the snapshots may contribute to those differences.
Our models predict warmer temperatures in the middle
  photosphere, and as a consequence, a less steep temperature gradient in
  layers where many of the spectral lines that play a role in solar chemical
  abundance determination form. This can have a large effect on the emerging
  radiation spectrum, given that the excitation or ionisation processes
  involved in spectral line formation are predominantly dependent on 
  that gradient. The
  results of this paper, namely the noticeable changes in the line spectra
  when going from HD to MHD and the excellent match that can be achieved in
  continuum and lines for the MHD models, together with further results in
  the literature like those of Papers \citetalias{Fabbian+2010} and
  \citetalias{Fabbian+2012} or those of \citet{Beck+2013,Beck+2015a}, which
  indicate an accurate average temperature stratification for our models with
  $\left< B_z \right>$ between $\sim 50$~G and $\sim 100$~G,
confirm the need to consider
  magnetoconvection processes when solving problems that have a sensitive
  dependence on the shape of spectral features. 

\ 

\ 

\section{Conclusions}\label{sec:conclusions}

In this work, we have studied the fit of various spectral features
obtained through our 3D MHD numerical models of convection to
observational constraints, finding, in general a very good agreement.
This is particularly the case for absolute continuum intensity
matching as well as for centre-to-limb variation in the wavelength
range $600$-$2400$ nm. The adopted opacities and the 3D {\it a
  posteriori} radiation transfer processing prove to be important for
a good match of observed solar continuum intensity from disc centre to
limb.  We have also shown that it is possible to achieve a good fit of
the difficult [O {\sc i}] $557$~nm and $630$~nm spectral features in
spite of the blends and peculiarities of these lines. A convection
model with the \citet{Asplund+2009} abundances [$\log \epsilon
(\textrm{C})_{\odot}=8.43$~dex and $\log \epsilon
(\textrm{O})_{\odot}=8.69$~dex] does not yield a satisfactory match,
neither for the HD nor for any of the MHD cases. A better fit can be
obtained by modifying the input abundances; for instance, a very good
fit is obtained for our MHD ``200~G'' model using abundance values
higher than those of \citet{Asplund+2009} by several centidex. Also
importantly, the presence of magnetic fields in the convection models
is seen to cause non-negligible changes in the line profiles
(especially in the [O \textsc{i}] $630$-nm case). This effect occurs
indirectly via the modification in the average temperature
stratification of the atmosphere in the line-forming layers.
The line asymmetries are also seen to be marginally
  different in the HD and MHD cases, so we cannot exclude the possibility
  that the differences in the average flow fields between the HD and MHD
  calculations play a role as well in modifying the line profile shapes. 
Adding to the results
in this paper those from previous publications on abundances (Paper
\citetalias{Fabbian+2010, Fabbian+2012}), or those dealing with
thermodynamic properties or polarization signals from our numerical
models \citep{Beck+2013, Beck+2015a}, we conclude that one should not
neglect the magnetic field of the convection cells even in quiet Sun
regions when aiming at accurate spectral modeling and chemical
abundance determination.

\acknowledgments We gratefully acknowledge financial support by the
Spanish Ministries of Research and Innovation and of Economy through
projects AYA2011-24808 and CSD2007-00050. We acknowledge the computing
time granted through the DEISA SolarAct and PRACE SunFlare projects
and the corresponding use of the HLRS (Stuttgart, Germany) and FZJ-JSC
JUROPA (J\"ulich, Germany) supercomputer installations, as well as the
computer resources and assistance provided at the MareNostrum
(BSC/CNS/RES, Spain) and LaPalma (IAC/RES, Spain) supercomputers. We
thank H. Uitenbroek for guidance with his RH code, and R. Rezaei,
C.~A.~R. Beck and N. Vitas for useful discussions on the solar
continuum intensity and related topics, as well as N. Shchukina and J.
Trujillo Bueno for conversations about their recent
  work. We also acknowledge T.~M.~D.  Pereira for clarifications on the 3D LTE line
formation code and opacity package employed in Paper
\citetalias{Pereira+2013a}. We have made use of the NIST ASD v5
available at http://physics.nist.gov/asd and of the first release
(2014-02-05) of the ``VALD3'' data, as available from the web
interface of the Vienna Atomic Line Database (VALD) project at
http://vald.astro.univie.ac.at/~vald3/php/vald.php. Finally,
  we thank the anonymous referee for the helpful comments.

\clearpage

\end{document}